\documentclass[twocolumn,aps]{revtex4}
\usepackage{graphicx}
\usepackage{bm}
\usepackage{amssymb} 

\begin{document}

\title{Two-Electron Linear Intersubband Light Absorption in a 
Biased Quantum
Well}

\author{J. Dai$^a$,  M. E. Raikh$^a$, and T. V. Shahbazyan$^b$}

\affiliation{
$^a$Department of Physics, University of Utah, Salt Lake City,
UT 84112\\
$^b$Department of Physics, Jackson State University, Jackson, MS
39217} 

\begin{abstract}
We point out a novel manifestation of many-body correlations in the
{\em linear} optical response of electrons confined in a quantum well.
Namely, we demonstrate that along with conventional absorption peak at
frequency $\omega$ close to intersubband energy $\Delta$, there exists
an additional peak at frequency $\hbar\omega \approx 2\Delta$. This
new peak is solely due to electron-electron interactions, and can be
understood as excitation of {\em two} electrons by a {\em single}
photon. The actual peak lineshape is comprised of a sharp feature, due
to excitation of {\em pairs} of intersubband plasmons, on top of a
broader band due to absorption by two single-particle excitations. The
two-plasmon contribution allows to infer intersubband plasmon
dispersion from linear absorption experiments.
\end{abstract}

\maketitle

{\noindent \it Introduction} --- Intersubband absorption of light in a
quantum well (QW) is studied theoretically and experimentally for more
than two decades. Original motivation for such a close attention to
this process was its crucial role in design of infrared
detectors~\cite{choi97}. Lately, the interest in intersubband
transitions is spurred by advances in fabrication of the quantum
cascade lasers~\cite{liu00}.

\begin{figure}
 \begin{center}
 \includegraphics[width=1.1in]{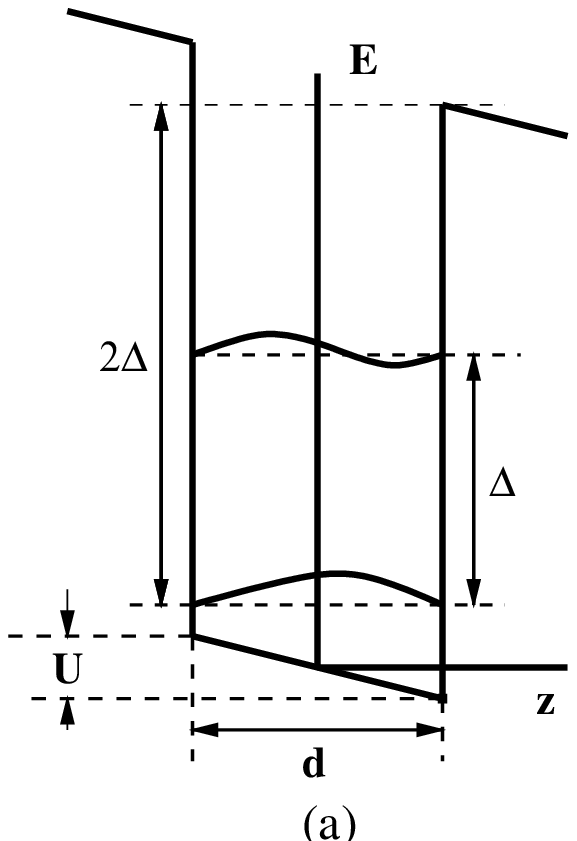}
 \hspace{10mm}
 \includegraphics[width=1.0in]{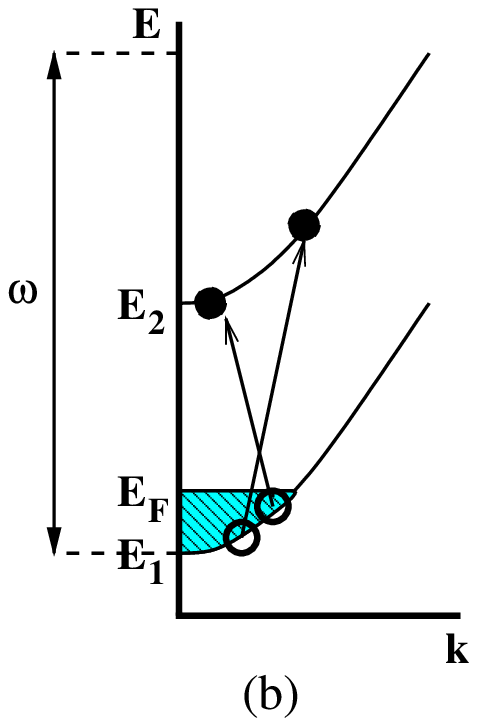}
 \end{center}
 \caption{(a) Two lowest size-quantization wave functions of a biased
 quantum well are shown schematically; for two-electron linear
 absorption, the incident photon energy
 must be close to $2\Delta$.
 (b) 
Two-electron intersubband
 absorption in the lowest order of the perturbation theory.}
 \label{fig:subbands}
 \end{figure}

Within a single-electron description and in the absence of
nonparabolicity, the intersubband absorption peak is infinitely narrow
and positioned precisely at $\hbar\omega=\Delta$, where
$\Delta=E_2-E_1$ is the intersubband separation (see
Fig.~\ref{fig:subbands}). As electron-electron interactions are
switched on, the adequate language for the description of
absorption becomes the excitations of intersubband plasmon (ISP) by
light polarized perpendicular to the QW plane.  Many-body origin of
the absorption manifests itself in a shift of peak position up from
$\hbar\omega=\Delta$ (depolarization shift), and a finite peak width
even at zero temperature and in the absence of disorder.  While the
shift has been understood long ago~\cite{ando77,ando82,brey89}, the
interaction-induced broadening of the absorption line still remains a
subject of debate. The peak lineshape was addressed in several recent
studies that employed various approximate many-body schemes, and
yielded lineshapes calculated numerically for particular sets of QW
parameters\cite{ullrich01,faleev02,li03,waldmuller04,pereira04,li04}.
The lack of analytical description (even in the limit of weak 
interactions) reflects the fact that the peak lineshape is governed by
very delicate correlations in 2D electron gas.
%
%

In the present paper we point out that, in addition to conventional
intersubband absorption, there exists a distinctive 
{\em interaction-induced} effect, namely, {\em two-particle linear
absorption} of light. Obviously, without interactions, one photon can
excite only {\em one} electron from the lower to the upper
subband. This is the case even in the presense of disorder that
violates momentum conservation. Our main point is that interactions
allow a photon with energy $\hbar\omega \approx 2\Delta$ to excite
simultaneously {\em two} intersubband excitations (single-particle or
collective).  As a result, the absorption peak is comprised of a sharp
feature originating from excitation of two {\em ISP}s
on top of a broader band 
of
two {\em single-particle excitations}.  We show that absorption of
a photon by two 
single-particle excitations can be captured perturbatively in the
interaction strength. By contrast, in order to describe two-plasmon
absorption, a nonperturbative approach is required. Such an approach
is developed in this paper.

A remarkable feature of two-particle absorption
is that the ISP dispersion $\omega_{pl}(q)$ at 
{\em finite} momenta can be inferred from the peak shape.  Indeed,
with two finite-momentum ISPs in the final state, the momentum
conservation can be respected even though the momentum of an incident
photon is negligible~\cite{koch04}.  Another distinctive feature of
the $2\Delta$ peak is its sensitivity to external bias, $U$, applied
across QW; 
for a symmetric QW, the $2\Delta$ peak emerges
only at finite $U$ and grows 
as $U^2$.

\begin{figure}
 \begin{center}
 \includegraphics[width=2.4in]{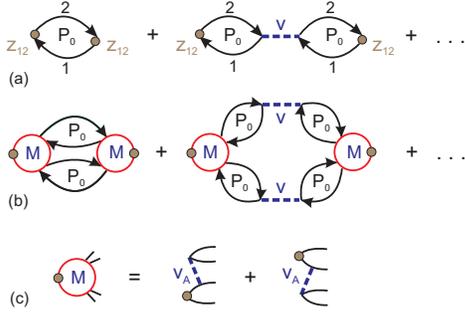}
 \end{center}
 \caption{Diagrammatic representation of single-particle absorption
   (a), two-particle absorption (b) and two-electron dipole matrix 
   element (c). $V_A$ stands for $V_{1222}$ and $V_{1211}$ in first and
   second digram, respectively.}
 \label{fig:diagram}
 \end{figure}

{\noindent \em Formalism} --- The proposed effect is most naturally
described in diagrammatic language.The conventional absorption is
represented by standard polarization bubble 
$\bigl[$see Fig.\ \ref{fig:diagram}(a)$\bigr]$. Dipole matrix elements 
$z_{12}=\int \varphi_1 z \varphi_2 dz$ in the vertices are responsible
for promotion of an electron from subband 1 to subbband 2 (see 
Fig.\ \ref{fig:subbands}), where $\varphi_1$, $\varphi_2$ are the
size-quantization wavefunctions.  Since the momentum of photon is
negligibly small, the intersubband polarization operator is simply
$P_0(\omega)=\frac{N}{\omega-\Delta}$, where $N=\frac{k_F^2}{2\pi}$ is
electron concentration and $k_F$ is the Fermi momentum. We assume that
the Fermi energy $E_F=\frac{\hbar^2k_F^2}{2m}$ is smaller than
$\Delta$ (here $m$ is the electron mass). As explained above, the
adequate picture of absorption is excitation of ISP (rather than
electron-hole pair).  Diagrammatically 
$\bigl[$see Fig.\ \ref{fig:diagram}(a)$\bigr]$ this is accomplished by
a standard RPA summation of polarization bubbles connected by
intersubband Coulomb matrix element, $V\equiv V_{1212}(0)$; we adopt
the standard definition of matrix elements 
\begin{eqnarray}
\label{Coulomb}
 V_{ijkl}(q)=
v_q \int dzdz'
 \varphi_{i}(z)\varphi_{j}(z')
 \varphi_{k}(z')\varphi_{l}(z)
 e^{-q\vert z-z'\vert},
\end{eqnarray}
where $v_q=\frac{2\pi e^{2}}{\kappa q}$ is the Fourier component of 2D
Coulomb potential and $\kappa$ is the dielectric constant.
For a rectangular well, we have
$V_{1212}(0)=\lambda_0/\nu$, where $\nu=m/\pi \hbar^2$ is the 2D
density of states, while the dimensionless parameter $\lambda_0$ is
given by
\begin{eqnarray}
\label{lambda}
\lambda_0=\frac{10}{9\pi} r_s \sqrt{6E_F/\Delta},
\end{eqnarray}
where $r_s=\sqrt{2}me^2/\kappa\hbar^2k_F$ is the interaction
parameter.  The RPA summation amounts to a replacement of $P_0$ by the
ISP propagator $\Pi=P_0(1-V_{1212}P_0)^{-1}$. It is important that at
$q=0$ the propagator $\Pi(\omega)=\frac{N}{\omega-\Delta-V_{1212}N}$
has the {\em same} form as $P_0(\omega)$ except for the depolarization
shift of the pole position \cite{ando77,ando82,brey89}. Thus, within
RPA, the {\em entire} oscillator strength is simply transferred to
ISP.
%
%

It is convenient to explain the novel absoption peak at $\omega\approx
2\Delta$ by building on analogy to Fig.\ \ref{fig:diagram}(a). The
corresponding diagrams are presented in 
Fig.\ \ref{fig:diagram}(b). The vertices $M_{\bf q}$ now stand for
matrix elements that transform a photon into {\em two} intersubband
electron-hole ({\em e-h}) pairs.  Each electron (hole) propagator in
Fig.\ \ref{fig:diagram}(a) is now replaced by {\em double} line (one
electron and one hole). All four lines begin (and end) at the same
vertex.

It is obvious that, without interactions, $M_{\bf q}$ is identically
zero. This is the case even in the presence of disorder and reflects
the orthogonality of eigenstates in QW. Our prime observation is that
interactions give rise to a {\em finite} $M_{\bf q}$. In the lowest
order in interactions, the diagram for $M_{\bf q}$ is shown in 
Fig.\ \ref{fig:diagram}(c).  The underlying virtual process can be
described as follows. A photon with energy $\omega\approx 2\Delta$
first creates an {\em e-h} pair by promoting an electron to the second
subband. The electron from this pair subsequently undergoes {\em
intrasubband} scattering, accompanied by excitation of 
{\em intersubband} {\em e-h} pair.  The propagator of this pair is
shown by the first double-line emerging from the vertex $M_{\bf q}$ in
Fig.\ \ref{fig:diagram}(b), while the second double-line corresponds to
intrasubband-scattered photoexcited electron and photoexcited hole.
Apparently, there is also a second contribution to $M_{\bf q}$,
originating from intrasubband scattering of the hole, as shown in
Fig.\ \ref{fig:diagram}(c).

We now turn to the higher-order diagram in Fig.\ \ref{fig:diagram}(b).
In higher orders in interactions, each double-line, that represented 
a propagating {\em e-h} pair in the lowest-order diagram, is now replaced
by ISP propagator.
Such a ``dressing'' is analogous to single-particle to collective
excitation transformation in the usual absorption [see 
Fig.\ \ref{fig:diagram}(a)]. Correspondingly, the absorption
coefficient has a general form
\begin{eqnarray}
\label{alpha2}
\alpha_2(\omega) \propto 
\omega \sum_{\bf q}|M_{\bf q}|^2\, J(\omega,q)\, ,
\end{eqnarray}
(we omitted a frequency-independent factor), where the joint spectral
function $J(\omega,q)$ is given by a convolution
\begin{eqnarray}
\label{joint}
J(\omega,q)=\int \frac{dE}{2\pi} \,
2 \, {\rm Im} \Pi(E,{\bf q})
\, 
2 \, {\rm Im} \Pi(\omega-E,{\bf q}),
\end{eqnarray}
of two ISP propagators. Energy dependence of $\Pi(E,{\bf q})$ comes
from the free-electron intersubband polarization,
\begin{eqnarray}
\label{P0}
P_0(E,{\bf q})= 2\int \frac{d{\bf{p}}}{(2\pi)^2} 
\frac{n_{1\bf{p}}}{E+\epsilon_{\bf p}-\Delta -\epsilon_{{\bf p}+{\bf q}}+i0}
\nonumber\\
= \frac{\nu}{2 \epsilon_{\bf q}}
\Biggl[
E - \Delta -\epsilon_{\bf q}
-\sqrt{(E -\Delta - \epsilon_{\bf q})^2-4E_F\epsilon_{\bf q}}
\Biggr].
 \end{eqnarray}
where $\epsilon_{\bf q} = \hbar^2q^2/2m$ is electron dispersion,
$n_{1{\bf p}}$ is the occupation of $n=1$ subband, and the factor 2
accounts for spin. From the pole position of $\Pi(E,{\bf q})$ at
$1=V_{1212}(q)P_0(\Omega_{pl},{\bf q})$, we obtain the ISP dispersion
law,
\begin{equation}
\label{dispersion}
  \hbar\Omega_{pl}(q) = \hbar\Omega_{pl}(0) + (\lambda_q-\lambda_0)E_F+
  (1+\lambda_q^{-1})\, \epsilon_{\bf q},
 \end{equation}
with $\hbar\Omega_{pl}(0)=\Delta + \lambda_0 E_F$ and $\lambda_q = \nu
V_{1212}(q)$, where $V_{1212}(q)$ is defined by Eq.\ (\ref{Coulomb}).
The bottom of the ISP band is shifted up by $\lambda_0E_F$ from the
intersubband separation $\Delta$. Note, that the full depolarization
shift includes also the Hartree renormalization of the intersubband
separation, which has the same order of $\lambda_0 E_F$.  For
parabolic QW this renormalization insures that $\Omega_{pl}(0)=\Delta$
in accordance with the Kohn theorem~\cite{brey89,brey90}.  For
relevant momenta $qd \lesssim 1$, the $q$-dependence of the matrix
elements, $V_{ijkl}$, can be neglected, which leads to quadratic ISP
dispersion: $\Omega_{pl}(q)-\Omega_{pl}(0)=q^2/2m_{pl}$, where
$m_{pl}=m\lambda_0/(1+\lambda_0)$ is the plasmon mass.  Remarkably,
for $E_F/\Delta \ll 1$, the plasmon is {\em much lighter} than the
elctron, $m_{pl}/m\sim \lambda_0\ll 1$.

The ISP contribution to $J$ comes from the region
$E-\Delta>2\sqrt{E_F\epsilon_{\bf q}} +\epsilon_{\bf q}$ of the
$(E,q)$ plane, where the plasmon is not Landau-damped.  Then, for the
plasmon propagator, $\Pi(E,q)$, one has $-2 {\rm Im}\Pi(E,q)=2\pi
A_q\delta\bigl[E-\hbar\Omega_{pl}(q)\bigr]$, with the oscillator
strength, $A_q$, given by
\begin{equation}
  \label{eq:oscillator}
  A_q =-\Biggl[V_{1212}(q)\frac{\partial P_0(E,q)}
  {\partial E}\Biggr]^{-1}_{E=\Omega_{pl}(q)}
  =
  \nu \biggl( E_F - \frac{\epsilon_q}{\lambda_q^2}\biggr) > 0.
\end{equation}
Using the fact that $qd \ll 1$, we can set 
$\lambda_q\approx \lambda_0$. Then the joint spectral
function(\ref{joint-plasmon}) acquires a simple form
\begin{equation}
  \label{joint-plasmon}
 J_{pl}
   = 2\pi \nu^2 
  \Bigl(E_F - \frac{\epsilon_q}{\lambda_0^2}\Bigr)^2
  \theta\Bigl(E_F - \frac{\epsilon_q}{\lambda_0^2}\Bigr)
  \delta\Bigl[\omega-2\Omega_{pl}(q)\Bigr],
\end{equation}
where the $\theta$-function restricts the momenta to the domain
where ISP is undamped.

The contribution Eq.\ (\ref{joint-plasmon}) comes from the poles of
$\Pi$ in the integrand of Eq.\ (\ref{joint}). For conventional
intersubband absorption, $\bigl[$Fig.\ \ref{fig:diagram}(a)$\bigr]$,
the pole contribution carries almost the entire oscillator
strength. By contrast, Eq.\ (\ref{joint-plasmon}) yields only an 
{\em additional} contribution to the joint spectral function, whereas
the {\em main} contribution comes from the 
numerators of 
$\Pi=P_0/(1-V_{1212}P_0)$ in
Eq.\ (\ref{joint}).  In calculating the latter, the denominators can
be set to $1$, i.e., ISPs 
are damped in the corresponding domain of the $(E,q)$
plane. In other words, the main contribution, 
\begin{eqnarray}
  \label{joint-electron}
 J_{ee}(\omega,q) = 
\int\frac{d{\bf k}_1 d{\bf k}_2}{(2\pi)^4}\, n_{1{\bf k}_1}n_{1{\bf
    k}_2}\,
\qquad  \qquad \qquad \qquad
\nonumber\\
\times
 8\pi \delta \bigl(\omega-2\Delta - \epsilon_{{\bf k}_1-{\bf q}} 
 -\epsilon_{{\bf k}_2+{\bf q}}
 +\epsilon_{{\bf k}_1} +\epsilon_{{\bf k}_2}\bigr),
\end{eqnarray}
is due to excitation of two {\em e-h} pairs rather than plasmons and, being
nonresonant, describes a broad plateau on which a sharp two-ISP
peak resides.

Let us turn to the form of the two-electron matrix element
$M_{\bf q}$.  Analytical expression, corresponding to the sum of the
diagrams in Fig. \ref{fig:diagram}(c), that describe possible
channels of two-electron excitation, has the form
\begin{eqnarray}
\label{dipole}
M_{\bf q}
=
2z_{12} \,
 \frac{V_{1222}(q)- V_{1211}(q)}
 {\Delta + \epsilon_{{\bf k}_1-{\bf q}}+\epsilon_{{\bf k}_2+{\bf q}}
   -\epsilon_{{\bf k}_1} - \epsilon_{{\bf k}_2}}.
\end{eqnarray}
Minus sign in Eq.\ (\ref{dipole}) originates from the difference in
energies of intermediate states for the two channels.  Note, that the
momenta of the final-state single-particle energies in the denominator
of Eq. (\ref{dipole}) are restricted by energy
conservation $\hbar\omega = 2\Delta + \epsilon_{{\bf k}_1-{\bf q}}
+\epsilon_{{\bf k}_2+{\bf q}} -\epsilon_{{\bf k}_1} 
-\epsilon_{{\bf k}_2}$. This relation ensures that the matrix element
Eq.\ (\ref{dipole}) depends only on the transferred momentum, $q$, so
that 
\begin{eqnarray}
\label{dipole2}
M_{q}
=
\frac{2z_{12}}{\omega-\Delta} \,
 \Bigl[V_{1222}(q)- V_{1211}(q)\Bigr].
\end{eqnarray}
%

{\noindent \em Absorption coefficient} --- It is convenient to express
the two-electron absorption coefficient, $\alpha_2(\omega)$, relative
to the single-electron intersubband absorption, $\alpha_1(\omega)$.
Namely, we introduce the ratio
\begin{equation}
\label{alpha-normalized}
\tilde{\alpha}_2(\omega)=\frac{\alpha_2(\omega)}
{\int d\omega^{\prime} \alpha_1(\omega^{\prime})}
=\frac{\alpha_2(\omega)}{2\pi \nu E_F\Delta z_{12}^2},
\end{equation}
so that the area under the peak $\tilde{\alpha}_2(\omega)$ is equal to
the ratio of the corresponding oscillator strengths.  To proceed
further, we note that the difference, $V_{1222}(q)-V_{1211}(q)$, in
the rhs of Eq.\ (\ref{dipole2}) has a general form 
$\frac{2\pi e^2d}{\kappa} f(qd)$, where $f(z)$ is a dimensionless
function. Then the ISP contribution to $\tilde\alpha_2$ can be
presented as 
%
\begin{equation}
\label{alpha-plasmon}
\tilde{\alpha}_2(\omega)
=\frac{4\omega E_F^2 r_s^2}
{(\omega - \Delta)^2}\frac{md^2}{\hbar^2\Delta} [f(0)]^2 \Phi (\omega),
\end{equation}
where 
$\Phi = \frac{2}{\pi \nu^3 E_F^2}
\int \frac{d{\bf q}}{(2\pi)^2}J_{pl}(\omega,q)$ is a dimensionless
function of $\omega$, and we assumed that $qd\ll 1$. For ISP
contribution this assumption is justified, since 
$qd\sim \lambda_0(k_Fd)$. Then, using  Eq.\ (\ref{joint-plasmon}), we
obtain 
\begin{eqnarray}\label{threshold}
\Phi_{pl} 
&& \approx 
\frac{\lambda_0}{1+\lambda_0}\,
\biggl[1 - \frac{\omega-2\Omega_{pl}(0)}{\delta\omega_{pl}}\biggr]^2,
 \end{eqnarray}
where the width, $\delta\omega_{pl}$, is given by
$\hbar\delta\omega_{pl} = 2\lambda_0(1+\lambda_0)E_F$.  Note that
$\Phi_{pl}$ is non-zero only in the interval 
$2\Omega_{pl}(0) < \omega <2\Omega_{pl}(0)+\delta\omega_{pl}$.  Since
this width $\delta\omega_{pl} \lesssim E_F$, we conclude that the
plasmon peak constitutes a sharp feature on top of a wider
two-electron band in $\tilde\alpha_2(\omega)$.
%
%
 \begin{figure}
 \centering
 \includegraphics[width=2.6in]{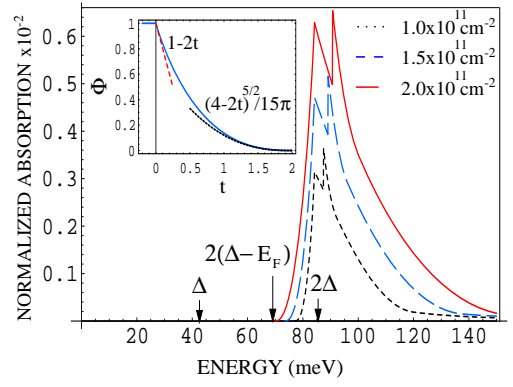}
\caption{Normalized two-electron absorption, $\Delta\tilde{\alpha}_2$,
 in $d=20$ nm GaAs QW is shown for applied bias $U=30$ meV and 
 electron concentrations $N=10^{11}$ cm$^{-2}$, $1.5\times 10^{11}$
 cm$^{-2}$, and 
 $2.0\times 10^{11}$ cm$^{-2}$. The right peak corresponds
 to two-plasmon 
 absorption. Arrow indicates the position of single-electron
 absorption peak at $\Delta$.
Inset: threshold behavior of
two-electron absorption calculated from Eq.\ (\protect\ref{phi-2e}).
}
\label{fig:absorp}
 \end{figure}
%

The two-electron contribution to $\tilde\alpha_2(\omega)$ can be
naturally divided into two frequency domains. First domain corresponds
to the photon energy $\omega-2\Delta\sim E_F$, as illustrated in 
Fig.\ \ref{fig:subbands}. 
In this domain we still have $qd \ll 1$, so that 
Eq.\ (\ref{alpha-plasmon}) applies but with $\Phi$ calculated using
$J_{ee}(\omega,q)$, defined by Eq.\ (\ref{joint-electron}). 
Upon performing the ${\bf q}$-integration, we obtain
\begin{eqnarray}
\label{phi-2e}
 \Phi(t) =
\frac{1}{\pi^2 k_F^4}
\int\limits_{k_1,k_2<k_F} d{\bf k}_1
d{\bf k}_2
\, \theta\left[\frac{\vert{\bf k}_1 - {\bf k}_2\vert^2}{2k_F^2} 
- t
\right],
\end{eqnarray}
with $t=(2\Delta-\hbar\omega)/E_F$.  The function $\Phi(t)$ evaluated
numerically is plotted in Fig.~\ref{fig:absorp} together with the
asymptotes $\Phi \approx (1-2t)$ for $t \ll 1$ and 
$\Phi\approx (4-2t)^{5/2}/15\pi$ for $(2-t) \ll 1$.  Singular points
$t=2$ and $t=0$ correspond to excitation of two electrons from the
bottom of the band and from the Fermi level, respectively. It follows
from Eqs.\ (\ref{threshold}) and (\ref{phi-2e}), that the ratio of
two-plasmon and two-electron contributions is equal to 
$\lambda_0(1+\lambda_0)^{-1}\ll 1$.

In the second domain, $\omega-2\Delta \gtrsim \Delta$, two excited
electrons have high energies. Thus, their momenta are almost opposite
to each other, both having the absolute value of
$q_{\omega}=\sqrt{2m(\hbar\omega-2\Delta)}/\hbar$.  Then the behavior
of $\tilde \alpha_2(\omega)$ in the high-frequency domain is given by
Eq.\ (\ref{alpha-plasmon}) with $[f(0)]^2\Phi$ replaced by
$[f(q_{\omega}d)]^2$. Thus, this behavior is determined by the actual
confinement potential profile. In the following we will consider the
most common example of a biased rectangular QW.

{\noindent \em Biased rectangular QW} ---
Without bias, due to the QW symmetry, we have $f(qd)=0$ for {\em all}
$q$.  In the presence of bias $U\lesssim \Delta$, which amounts to the
perturbation $Uz/d$ of confining potential, the
size-quantization wave-functions in Eq.\ (\ref{Coulomb}) acquire
symmetry-breaking corrections, $\varphi_1(z)=\varphi_1^{(0)}(z)
-\left(U z_{12}/\Delta d\right)\varphi_2^{(0)}(z)$, and
$\varphi_2(z)=\varphi_2^{(0)}(z) +\left(U z_{12}/\Delta
d\right)\varphi_1^{(0)}(z)$, where $\varphi_1^{(0)}$,
$\varphi_2^{(0)}$ are the size-quantization wave functions at $U=0$,
while the subband separation remains $\Delta=\frac{3\hbar^2\pi^2}{2md^2}$.
Then a straightforward calculation yields 
$f(qd)= \frac{Uz_{12}}{\Delta d}F(qd)$, where the function $F(s)$ can
be expressed analytically and has the following behavior:
$F(0)=595/144\pi^2$; $F(s)|_{s\gg 1}\approx 6/s^2$.  Thus, the
large-$\omega$ behavior of $\tilde\alpha_2(\omega)$ is the following:
$\alpha_2(\omega)\propto\omega/(\omega-\Delta)^2(\omega-2\Delta)^2$.
This slow decay is illustrated in Fig.~\ref{fig:absorp}, where 
$\tilde \alpha_2(\omega)$, caluclated numerically for realistic parameters
of quantum well, is plotted for three different electron
concentrations. The relative oscillator strength for two-electron
absorption is given by 
\begin{equation}
\label{net}
\int d\omega \tilde\alpha_2(\omega)=C\biggl(\frac{r_sE_FU}{\Delta^2}\biggr)^2.
\end{equation}
Using the explicit form of $F(s)$ we have numerically calculated the
coefficient $C$ in Eq. (\ref{net}) to be $8.2\times 10^{-2}$. 

{\noindent \em Concluding remarks} --- 
The lineshape of new peak is governed 
by three energy scales: sharp rise within the energy interval 
$2\Delta -2E_F <\hbar \omega < 2\Delta$, slow decay for 
$\hbar \omega > 2\Delta$, and a sharp two-ISP peak of
the width $\sim r_sE_F^{3/2}/\Delta^{1/2}$ on the top of the
two-electron band.   
This difference in scales justifies the fact that we have disregarded
the process of excitation of one ISP and one electron [this
process is also captured by Eq.\ (\ref{joint})]. The unusually small
ISP contribution can be traced to the smallness of the ISP
effectve mass. 

In general, the fact that a single photon can cause double ionization
of an interacting system with a {\em discrete} spectrum (such as
helium atom \cite{helium}) is known for almost four decades.
A remarkable feature of quantum wells is that a sharp
interaction-induced peak emerges in a system with {\em continuous}
electron spectrum\cite{homogeneous}. The necessary condition for
observation of the 
two-electron absorption is that quantum well must be deep enough,
namely deeper than $7\Delta/3$. Otherwise, two-electron absorption
will result in photoionization. 
The above condition is always satisfied for thick quantum
wells. However, increasing the thickness has a side effect that the
number of size-quatization levels in the well inceareses, thus
complicating the analysis of the absorption spectra.
Note finally, that although we
considered the simplest model of homogeneous electron gas at zero
temperature with two parabolic subbands, our theory can be easily 
generalized to the realistic situations~
\cite{nikonov97,perera98,pereira104,larrabee03}.
Important is that nonparabolicity, finite temperature, disorder,
etc. are not expected to suppress the two-electron absorption.

M.E.R. acknowledges support by NSF under Grant No. DMR-0503172. T.V.S.
acknowledges support by NSF under Grant No. DMR-0305557 
and by ARL under Grant No. DAAD19-01-2-0014.

\end{document}